\documentclass[10pt]{article}
\usepackage{amsmath}
\usepackage{amssymb}
\usepackage{graphicx}

\def\aa{A\&A }

\begin{document}

\setcounter{figure}{0}
\setcounter{table}{0}
\setcounter{footnote}{0}
\setcounter{equation}{0}

\vspace*{0.5cm}

\noindent {\Large GRAVITATIONAL REDSHIFT EXPERIMENT WITH THE SPACE\\
RADIO TELESCOPE RADIOASTRON}
\vspace*{0.7cm}

\noindent\hspace*{1.5cm}D. LITVINOV$^1$, N. BARTEL$^2$, K. BELOUSOV$^4$, M. BIETENHOLZ$^3$, A. BIRIUKOV$^4$,\\ 
\noindent\hspace*{1.5cm}A.~FIONOV$^1$,
A. GUSEV$^1$, V. KAUTS$^{4,5}$,
A. KOVALENKO$^6$, V. KULAGIN$^1$, N. PORAIKO$^1$,
\noindent\hspace*{1.5cm}V.~RUDENKO$^1$\\
\noindent\hspace*{1.5cm}$^1$ Lomonosov Moscow State University, Sternberg Astronomical Institute, Universitetsky pr.~13,
\noindent\hspace*{1.5cm}119991 Moscow, Russia, e-mail: litvirq@yandex.ru\\
\noindent\hspace*{1.5cm}$^2$ York University, Toronto, Ontario M3J 1P3, Canada\\
\noindent\hspace*{1.5cm}$^3$ Hartebeesthoek Radio Observatory, P.O. Box 443, Krugersdorp 1740, South Africa\\
\noindent\hspace*{1.5cm}$^4$ Astro Space Center of the Lebedev Physical Institute, Profsoyuznaya 84/32, 117997 Moscow, 
\noindent\hspace*{1.5cm}Russia\\
\noindent\hspace*{1.5cm}$^5$ Bauman Moscow State Technical University, 2-ya
Baumanskaya 5, 105005 Moscow, Russia\\
\noindent\hspace*{1.5cm}$^5$ Pushchino Radio Astronomy Observatory, 142290 Pushchino, Russia\\

\vspace*{0.5cm}

\noindent {\large ABSTRACT.} A unique test of general relativity is possible with the space radio telescope RadioAstron. The
ultra-stable on-board hydrogen maser frequency standard and the highly eccentric orbit make
RadioAstron an ideal instrument for probing the gravitational redshift effect. Large
gravitational potential variation, occurring on the time scale of $\sim$24 hr, causes large variation
of the on-board H-maser clock rate, which can be detected via comparison with frequency
standards installed at various ground radio astronomical observatories. The experiment
requires specific on-board hardware operating modes and support from ground radio
telescopes capable of tracking the spacecraft continuously and equipped with 8.4 or 15 GHz receivers. Our preliminary estimates show that $\sim$30 hr of the space radio telescope's observational time are required to reach $\sim 2\times10^{-5}$ accuracy in the test, which would constitute a factor of 10 improvement over the currently achieved best result.

\vspace*{1cm}

\noindent {\large 1. INTRODUCTION}

\smallskip

According to Einstein's principle of equivalence an electromagnetic wave propagating in a region of space where the gravitational potential is not
constant experiences a gravitational frequency shift, $\Delta f_\mathrm{grav}$, proportional to the
gravitational potential difference between the measurement points, $\Delta U$, and the frequency, $f$, of the wave:
\begin{equation}
   {\Delta f_\mathrm{grav} \over f} = { \Delta U \over c^{2} },
\label{eq:main}  
\end{equation}
where $c$ is the speed of light (Misner et al. 1973). Any violation of Eq. \eqref{eq:main} in an experiment with
two identical atomic frequency standards can be parameterized
in the following way:
\begin{equation}
{\Delta f_\mathrm{grav} \over f} = {\Delta U \over
c^2} (1+\varepsilon),
\end{equation}
where the violation parameter, $\varepsilon$, may depend on element composition of the gravitational field sources and on the kind of frequency standards. It is generally agreed that the best test of Eq. \eqref{eq:main}
to date was performed in the suborbital Gravity~Probe~A (GP-A) experiment, which measured $\varepsilon=(0.05\pm1.4)\times10^{-4}$ 
for two hydrogen masers (Vessot et al. 1980).
A similar experiment with RadioAstron, benefitting from a more stable hydrogen
maser (H-maser) and longer data acquisition, could tentatively measure $\varepsilon$
with an accuracy of $\delta\varepsilon \sim 2\times10^{-5}$. Below we outline two approaches to the anticipated experiment and give an
account of the technical tests made for it.

\vspace*{0.7cm}

\noindent {\large 2. OUTLINE OF THE EXPERIMENT}

\smallskip

In the gravitational redshift experiment with RadioAstron we use microwave
 radio links to monitor the
 redshifted frequency  of the satellite's on-board H-maser as it moves in the regions with different gravitational potential. The satellite radio 
payload includes two transmitters at 8.4 and 15 GHz and a 7.2 GHz receiver.
The transmitters can be fed with
a signal phase-locked either to the on-board H-maser, 
the 7.2 GHz uplink or a specific mixture of the two (see below). Measuring
the frequency of a one-way satellite downlink signal at a ground station we see it shifted by (Vessot \& Levine 1979):
\begin{equation}
\Delta f =
f \left(
 - {\dot D \over c} 
- {v_\mathrm{s}^2 - v_\mathrm{e}^2 \over 2 c^2}
+ { (\mathbf v_\mathrm{s} \cdot \mathbf n)^2
- (\mathbf v_\mathrm{e} \cdot \mathbf n) \cdot (\mathbf v_\mathrm{s} \cdot \mathbf n)
  \over c^2 }
\right)
+ {\Delta f_\mathrm{grav} }%\over f}
+ {\Delta f_\mathrm{ion} }%\over f} 
+ {\Delta f_\mathrm{trop} }%\over f}
+ {\Delta f_0 }%\over f}
+ O\left({v\over c}\right)^3,
\label{eq:one-way}
\end{equation}
where  $\dot D$ is the radial velocity of the
spacecraft relative to the ground station,  $\mathbf v_\mathrm{s}$ and $\mathbf v_\mathrm{e}$ are the velocities of the spacecraft and
the ground station, $\mathbf n$ is a unit vector in the direction opposite to that of signal propagation,   $\Delta f_\mathrm{grav}$ is the
gravitational redshift, $\Delta f_\mathrm{ion}$  and $\Delta f_\mathrm{trop}$ are the ionospheric and tropospheric shifts,  $\Delta f_0$ is an unknown frequency offset
between the ground-based and space-borne H-masers and each quantity is referred to the geocentric inertial reference
frame. Terms of $O\left({v\over c}\right)^3$
need to be taken into account only if aiming for an experiment accuracy of
$\delta\varepsilon\lesssim10^{-6}$
(Salomon et al. 2001).

The value of  $\Delta f_0$ could
be relatively large for H-masers due to their low intrinsic accuracy. For RadioAstron's H-maser ${\Delta f_0 / f} \sim 10^{-11}
$, which makes it impossible to  experimentally determine the total value of the gravitational redshift effect $\Delta U/c^2 \sim7\times10^{-10}$ with an
accuracy higher than $\sim$$10^{-2}$. However, since the rate
of change, or drift, of $\Delta f_0$   is typically small ($\sim$$1\times10^{-15}$ per day for RadioAstron), the relatively large value of $\Delta f_0$ does not prohibit us from conducting a high-accuracy experiment as long as
only the variation, but not the total value, of the gravitational redshift effect is to be determined. Then the fundamental limit to the accuracy, $\delta\varepsilon$, 
is set by the available gravitational potential variation along the orbit, the frequency standard's instability
and
its drift.
For RadioAstron this theoretical limit is $2\times10^{-6}$ if the experiments
are performed in the periods of the lowest perigee height  $\sim$1,000~km. 

\begin{figure}[h]
\begin{center}
\includegraphics[scale=.6]{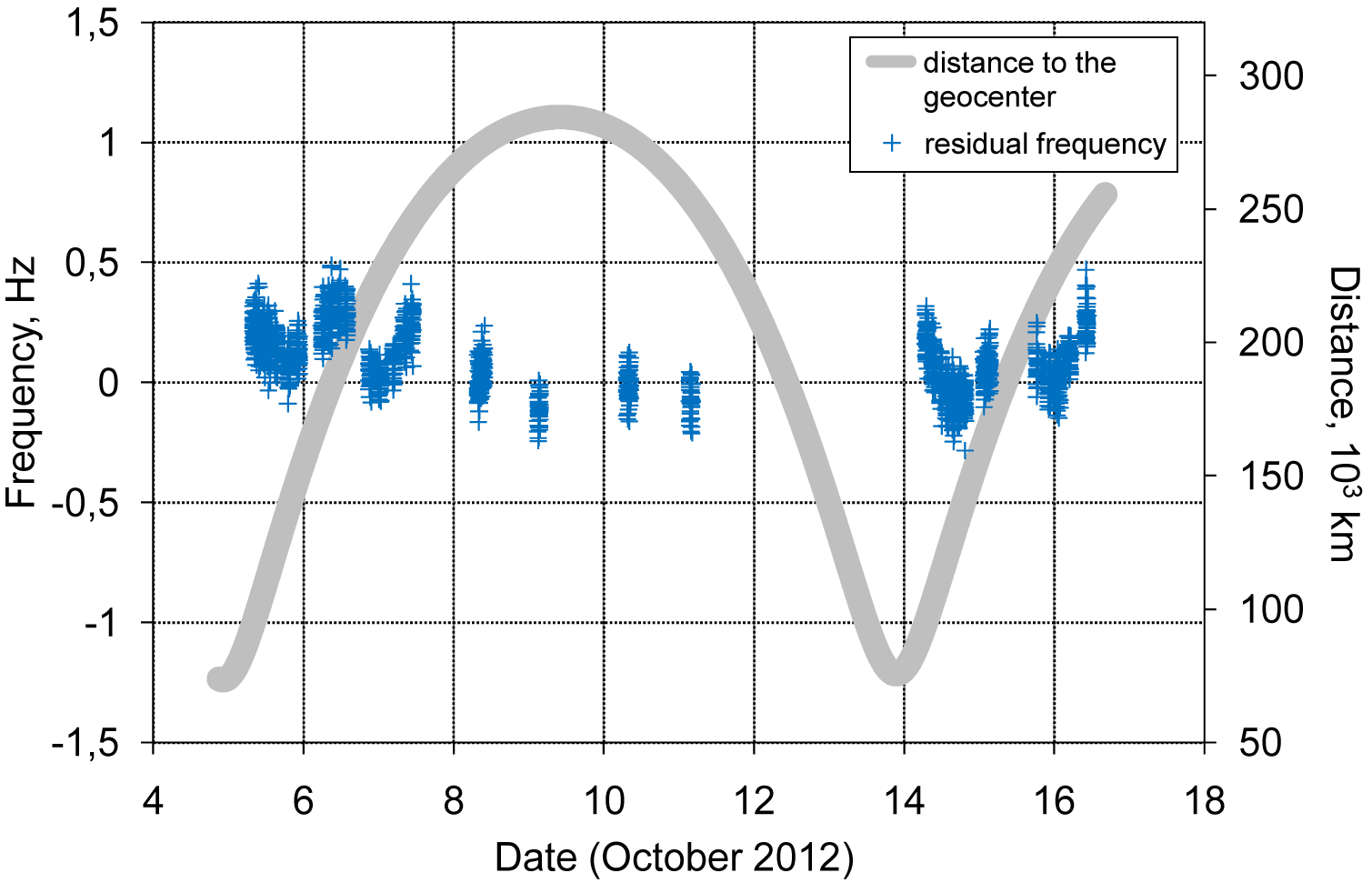}
\caption{Frequency residuals (observed--predicted) as a function
of epoch for the 8.4 GHz link.
RMS of residuals: 0.18 Hz; gravitational redshift $\Delta f_\mathrm{grav}$: 5 to 6 Hz (not plotted).}
\end{center}
\end{figure}

The principal source of error, when using Eq. (3) directly, is not the on-board H-maser performance
but the spacecraft radial velocity uncertainty  $\delta\dot D\sim 1 \;\mbox{mm/s},$
which sets the limit to the experiment accuracy $\delta\varepsilon\sim
3\%$ (Fig. 1).
%\ref{fig:residuals}).
Obviously, since the Doppler term cannot be determined
sufficiently accurately, the best would be to eliminate it completely
from the analysed signal. This is indeed possible if two kinds of radio links
are available, a one-way downlink, synchronized to the on-board H-maser, and
a two-way phase-locked loop (PLL), synchronized to the ground H-maser. 
The 1st-order Doppler shift of the two-way link is twice that 
of the  one-way downlink, but the gravitational frequency shift is zero. The signals of these two links
can be combined by
a radio engineering scheme, first used in GP-A, so that its output 
fully retains
the gravitational contribution but eliminates the
 1st-order Doppler term.

\begin{figure}[h]
\begin{center}
\begin{tabular}{ccc}
\includegraphics[scale=.29]{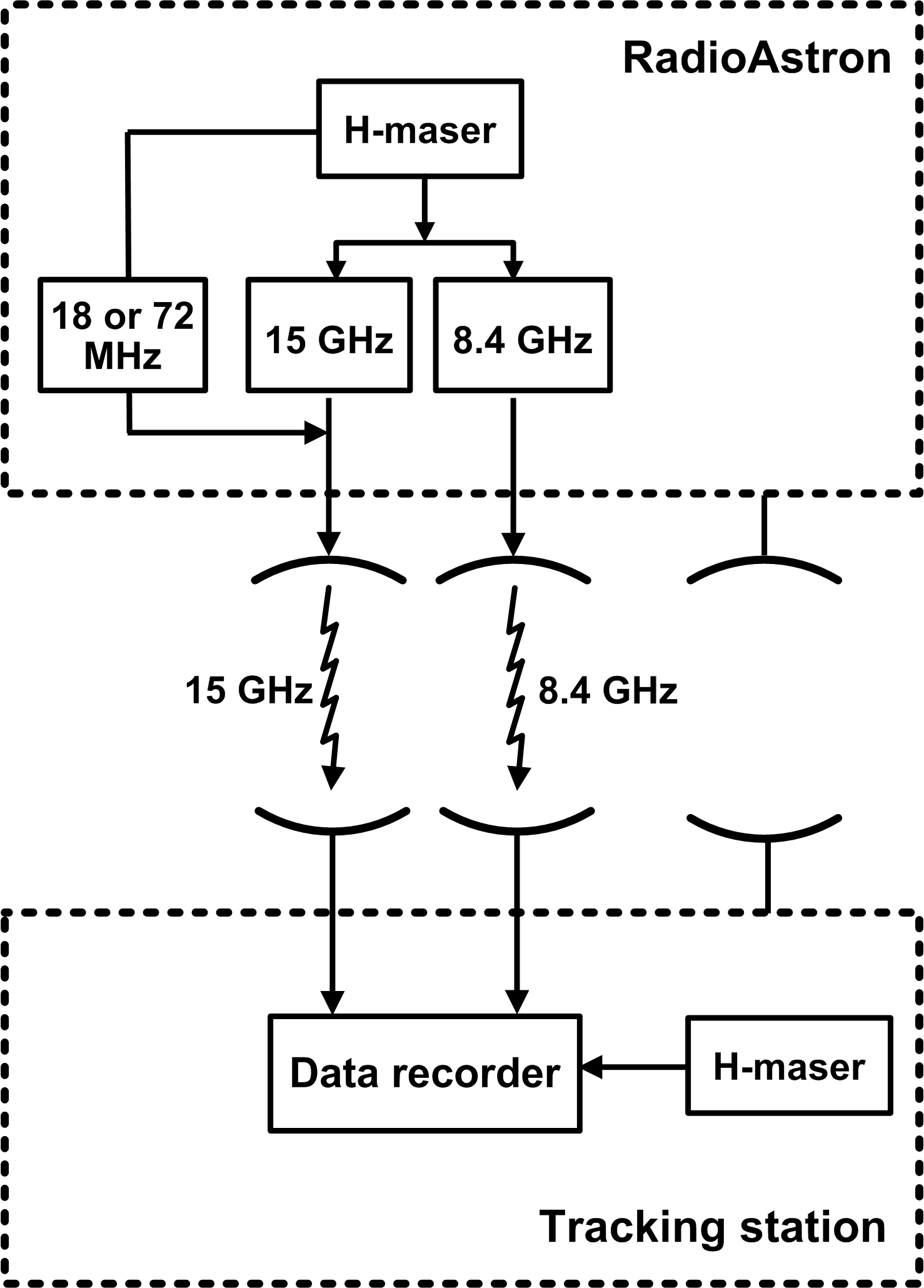}\quad\quad
&
\includegraphics[scale=.29]{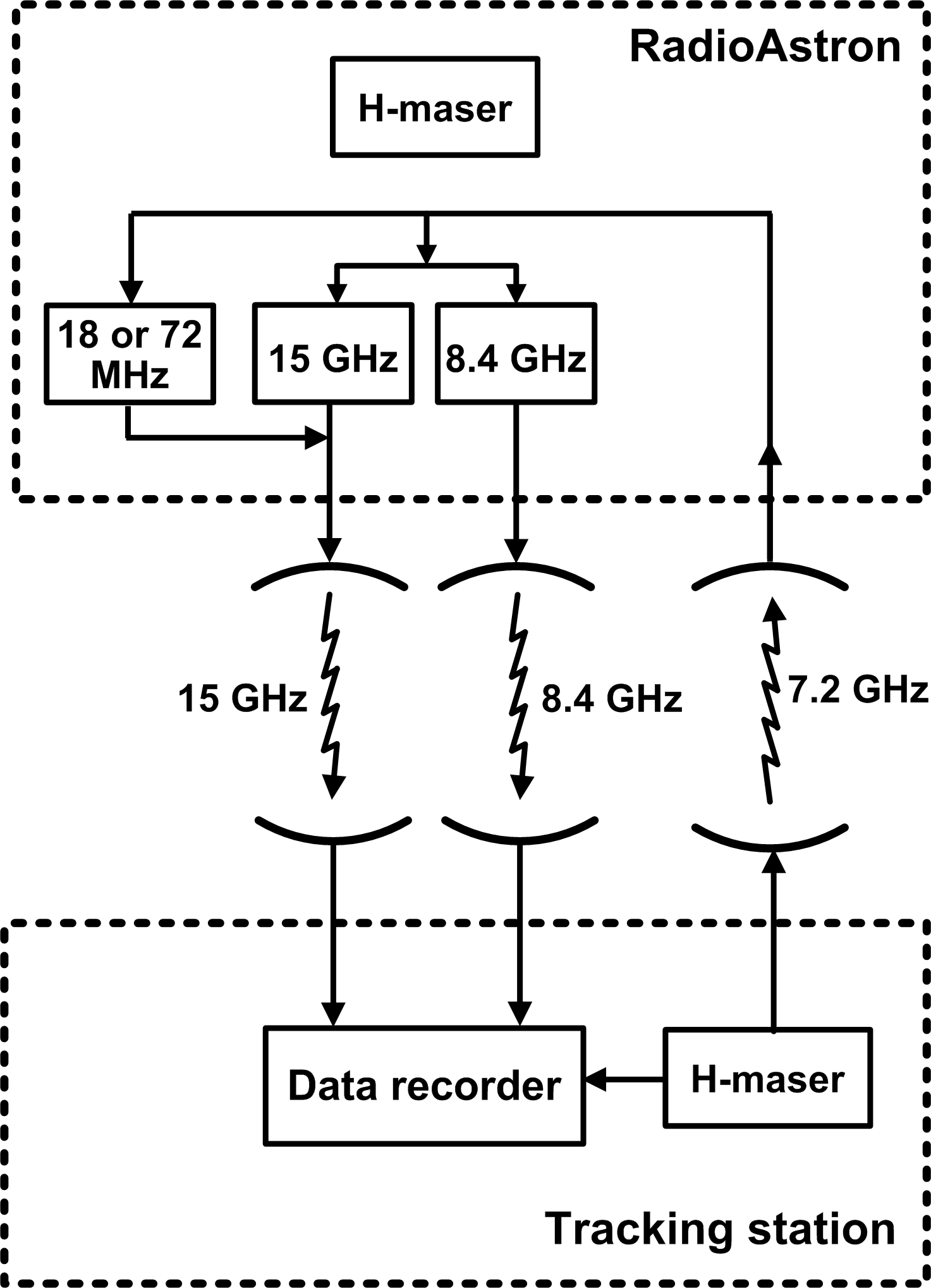}\quad\quad
&
\includegraphics[scale=.29]{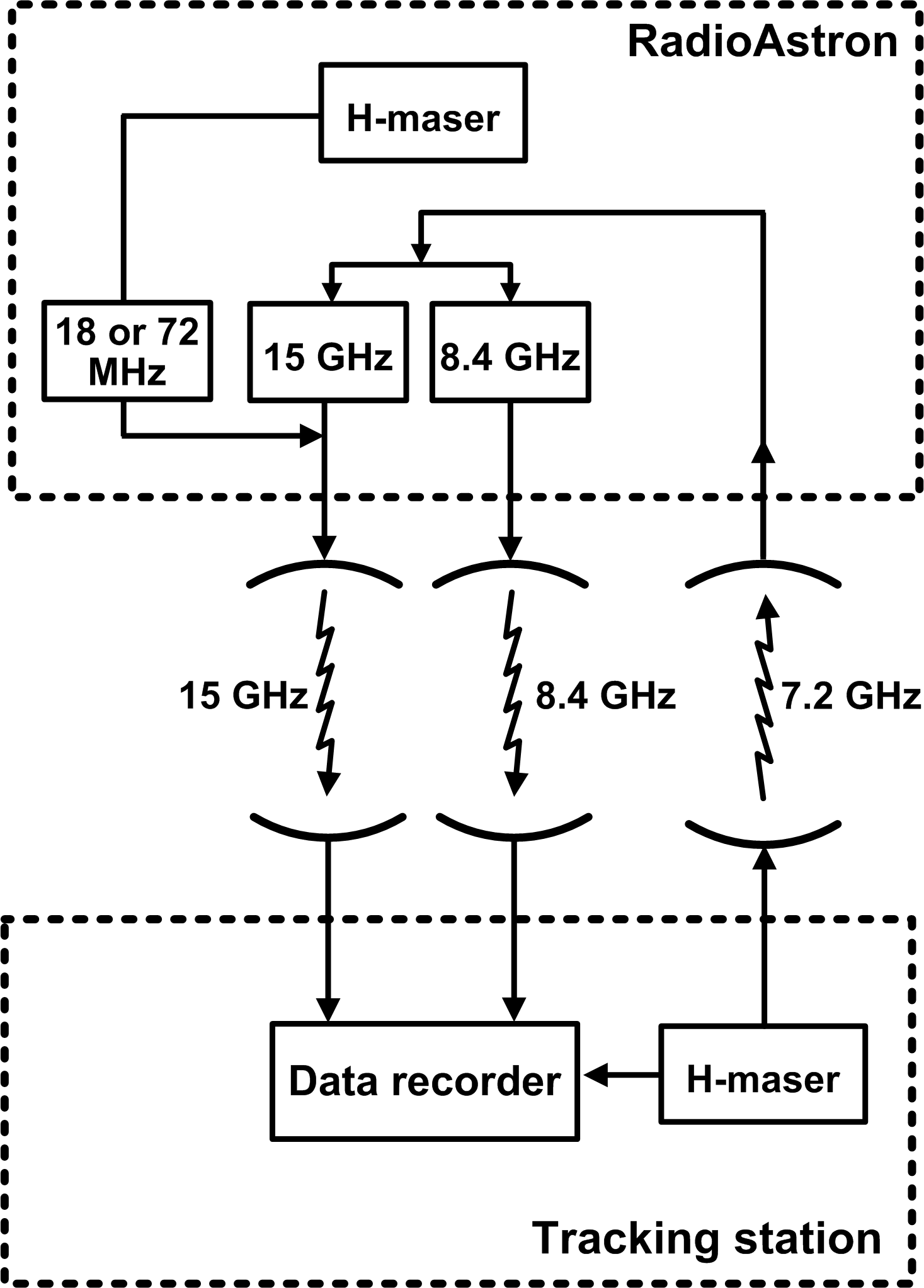}
\\
\vspace{0.2cm}
a) & b) & c)
\end{tabular}
\caption{On-board hardware  synchronization modes: a) ``H-Maser''; b) ``Coherent''; c) ``Semi-Coherent''. Note that the 8.4 GHz tone and the carrier of the 15 GHz data link cannot be synchronized independently. }
\end{center}
\end{figure}

For RadioAstron the GP-A compensation scheme 
is not directly applicable, because 1- and 2-way carrier frequency measurements (Fig. 2) cannot be performed simultaneously.
%\ref{fig:modes}).
Nevertheless, two modified versions of the Doppler compensation scheme are possible,
both of which rely on spacecraft tracking by ground
radio telescopes equipped with 8.4 or 15~GHz receivers (Duev et al. 2012).
The first option requires
 switching back and forth between the 1-way  (``H-maser'') and 2-way (``Coherent'') modes of  operation (Fig.~2a, b). Interleaving
the two synchronization modes results in two sets of gapped 1-way
and 2-way 
frequency measurements, which, after interpolation, allow for direct application
of the original GP-A 1st-order Doppler compensation scheme. The approach with interleaved measurements
does not rely on any features
of the signal spectrum, and thus can be realized with telescopes equipped
with any type of receiver (8.4 or 15 GHz).

The second approach to Doppler compensation involves
recording the 15~GHz  data link signal in the  ``Semi-Coherent'' mode of the on-board scientific and radio equipment (Fig.~2c). In this mode the  7.2~GHz
uplink tone, the 8.4 GHz downlink tone and the 15 GHz data downlink
carrier are phase-locked to the ground H-maser signal, while the modulation frequency
  of the data downlink is phase-locked to the on-board H-maser signal.
This approach also depends on the broadband ($\sim$1~GHz) nature of the QPSK-modulated
15~GHz signal and the
possibility of turning its spectrum into a comb-like form by  transmitting
a predefined periodic data sequence (Fig.~3). 
It was shown by Biriukov et al. (2014) that
different subtones of the resulting spectrum act like separate
links of the GP-A scheme and can be organized in software postprocessing into a  combination,
which is free from the 1st-order Doppler
and tropospheric noise terms (the ionospheric term persists).

 %\ref{fig:test-2}).
\begin{figure}[h]
\begin{center}
\includegraphics[scale=1.2]{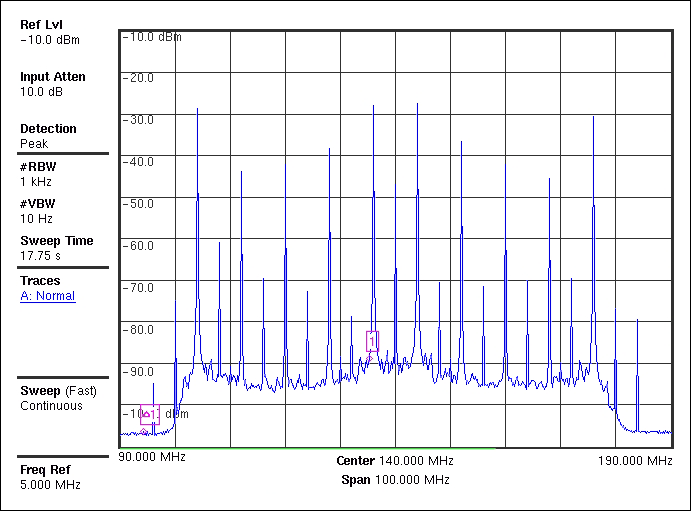}
\caption{15 GHz datalink signal spectrum in the ``Test-2'' 72 MHz mode of the on-board formatter.}
\end{center}
\end{figure}

Since in Europe only the Effelsberg telescope is equipped with a 15 GHz receiver, most
experiments supported by the RadioAstron mission's
Pushchino tracking station would
use the first approach to Doppler compensation. By contrast, experiments supported by the Green Bank tracking station could use any of the two approaches since the GBT and all VLBA antennas are
equipped with 8.4 and 15~GHz receivers and are capable of continuous spacecraft
tracking (however, only Hn, NL and, of course, the GBT are located sufficiently close to the Green Bank
tracking station to be able to observe RadioAstron during low perigee sessions).
A single experiment would be made in two 1-hr sessions, one close to perigee and another
close to apogee. The currently predicted RadioAstron orbit allows
for 10 to 15 experiments in 2015 to 2016 with a modulation of the gravitational
potential along the orbit of   ${\Delta U /
c^2}\sim3\times10^{-10}$ and   1 to 3 radio telescopes  tracking the satellite. With preliminary values for the Allan deviation of $\sim 3\times 10^{-14}$ at 1,000~s for the 1- and 2-way modes, the accuracy of the experiment could be as high as 
\begin{equation}
\delta\varepsilon\sim 2\times10^{-5}.
\label{eq:estimated-accuracy}
\end{equation}

\vspace*{0.7cm}
 
\noindent {\large 3. PRESENT STATUS OF THE EXPERIMENT}

\smallskip

Currently the experiment is in its testing phase. Up to now we have checked
the operability of the required on-board hardware modes and performed a series of recordings of the satellite downlink signals using regular VLBI equipment at the RadioAstron mission's  tracking station in Pushchino. The recovered signal
frequencies show good agreement with ordinary frequency measurements performed
at the tracking station as part of the mission support (Fig. 4).
\begin{figure}[h]
\begin{center}
\includegraphics[scale=0.29]{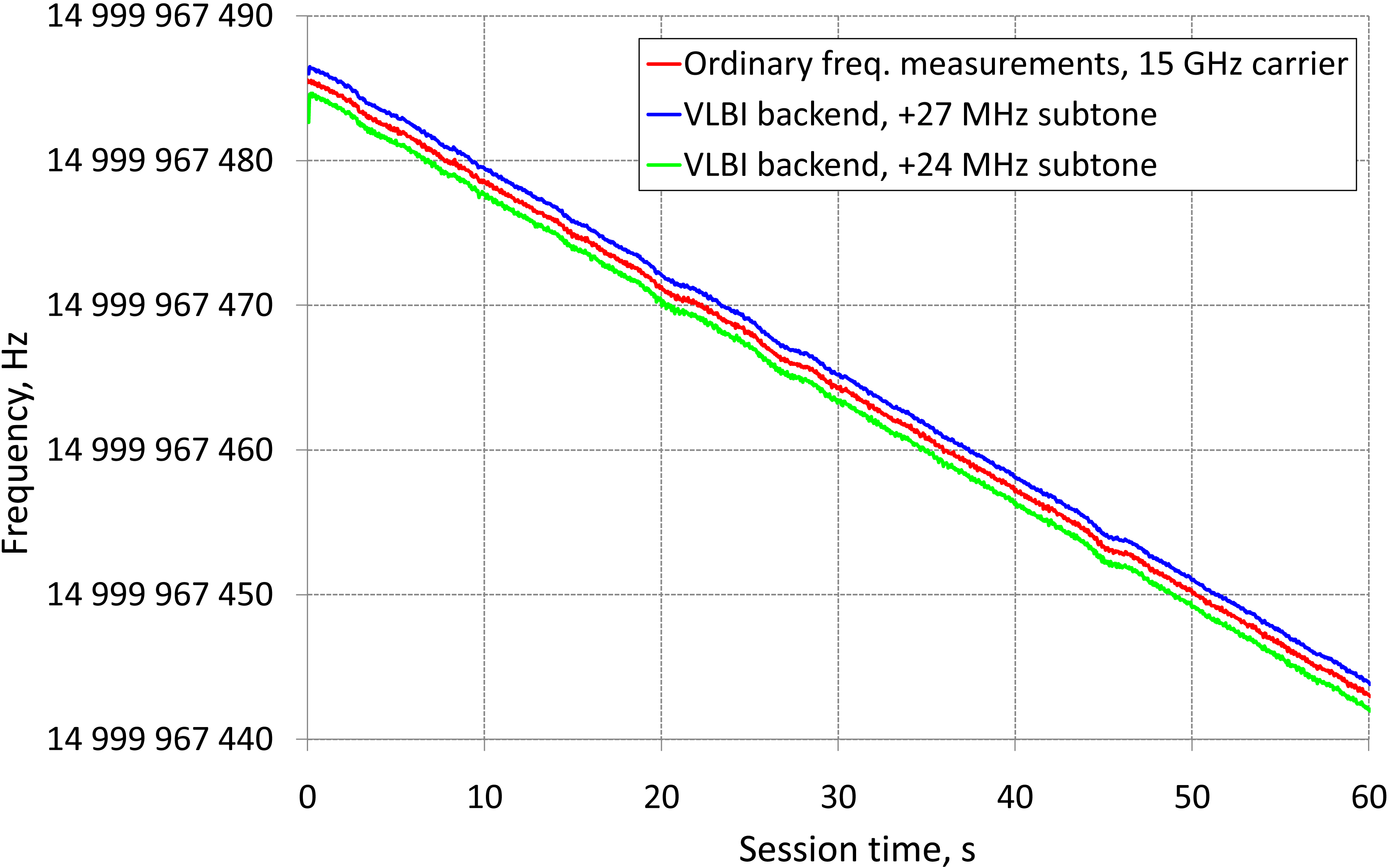}
\caption{Frequency measurements of the 15 GHz signal subtones in the ``Test-2'' 18 MHz mode made at the Pushchino tracking station, 2014/08/31 08:20:00 UTC. The carrier frequencies
were measured using standard tracking station equipment, the subtone frequencies were recovered from a 2-bit quantization 32 MHz bandwidth recording made by a  VLBI backend.
Subtone frequencies are offset by $\sim$24 and $\sim$27~MHz for easier comparison with  the carrier frequency measurements. 
}
\end{center}
\end{figure}
Their stability (Allan deviation of $6\times10^{-14}$ at 1,000~s) is lower than required
to achieve $\delta\varepsilon \lesssim 2 \times 10^{-5}$ but in accord with previous
satellite tracking experiments at Pushchino. The recordings obtained from the first RadioAstron tracking test in the 2-way mode by a number of EVN and Asian telescopes exhibit at least 2 times better signal stability  and give reason to believe that the above accuracy of the gravitational redshift test can be achieved.    

\vspace*{0.5cm}
 
The "RadioAstron" project is led by the Astro
Space Center of the Lebedev Physical Institute of
the Russian Academy of Sciences and the Lavochkin
Scientific and Production Association under a contract with the Russian Federal Space Agency, in collaboration with partner organizations in Russia and
other countries.

\vspace*{0.7cm}

\noindent {\large 4. REFERENCES}
% Please type the reference as follows
% Name Initial, year, "title", journal, vol. , pp. x-x.
%
% Examples:
%
% Author1, N., Author2, N., 2000, ``Title of the paper'', 
% \aa 111, pp. 111--222.
%
% Author2, N., Author3, N., 2003, ``Title of the paper'',
% \jgr (Solid Earth), 111(B5), doi: 10.1000/2002JB001111.
%
% PLEASE DO NOT USE ANY SPECIAL FONTS 
% (no italics, no boldface, etc.)
%
{

\leftskip=5mm
\parindent=-5mm

\smallskip

Biriukov A. V. et al., 2014, ``Gravitational Redshift Test with the Space
Radio Telescope RadioAstron'', Astron. Rep. 58, pp.783-795.

Duev D. et al., 2012, ``Spacecraft VLBI and Doppler tracking: algorithms
and implementation'', \aa 541, A43.

Misner, C., Thorne, K., Wheeler, J., Gravitation, San Francisco: Freeman, 1973.

Salomon et al., 2001, ``Cold atoms in space and atomic clocks: ACES'', C.
R. Acad. Sci. Paris, t. 2, S\'erie IV, p. 1313-1330

Vessot, R. F. C. et al., 1980, ``Test of Relativistic Gravitation with a Space-Borne Hydrogen Maser'', Phys. Rev. Let., 45, pp. 2081--2084.

Vessot, R. F. C., Levine M., 1979, ``A Test of the Equivalence Principle Using a Space-Borne Clock'', Gen. Rel. Grav., 10, 181-204.

}

\end{document}